\documentstyle[jkas2]{article}

\runningauthor{PARK AND PARK}
\runningtitle{Simulation of CMB Polarization for AMiBA}
\beginpage{1}
\endpage{7}

\def\uK{\mu {\rm K}}
\def\um{\mu {\rm m}}
\def\deg{^{\circ}}
\def\min{^{\prime}}
\def\fdg{\hbox{$.\!\!^\circ$}}

\def\btheta{\mbox{\boldmath $\theta$}}
\def\bhmap{\mbox{\boldmath $h$}}
\def\bmmap{\mbox{\boldmath $m$}}
\def\bu{{\bf u}}

\def\bx{{\bf x}}

\begin{document}

\title{Simulation of Cosmic Microwave Background Polarization Fields \\
       for AMiBA Experiment} 

\author{CHAN-GYUNG PARK AND CHANGBOM PARK}

\address{Astronomy Program, School of Earth and Environmental Sciences, 
       Seoul National University, 151-742 Korea\\
{\it E-mail: parkc@astro.snu.ac.kr and cbp@astro.snu.ac.kr}}

\address{\normalsize{\it (Received Mar. ??, 2002; Accepted ???. ??, 2002)}}

\abstract{
We have made a topological study of cosmic microwave background (CMB)
polarization maps by simulating the AMiBA experiment results.
A $\Lambda$CDM CMB sky is adopted to make mock interferometric observations
designed for the AMiBA experiment. CMB polarization fields are reconstructed
from the AMiBA mock visibility data using the maximum entropy method.
We have also considered effects of Galactic foregrounds on the CMB polarization
fields. The genus statistic is calculated from the simulated $Q$ and $U$
polarization maps, where $Q$ and $U$ are Stokes parameters.
Our study shows that the Galactic foreground emission, even at low Galactic
latitude, is expected to have small effects on the CMB polarization field.
Increasing survey area and integration time is essential to detect non-Gaussian
signals of cosmological origin through genus measurement.}

\maketitle

\section{INTRODUCTION}

The cosmic microwave background (CMB) polarization can provide useful
information in determining cosmological parameters that are only weakly
constrained by the CMB temperature anisotropy alone, such as the epoch
of reionization or the presence of tensor perturbations.
The Array for Microwave Background Anisotropy (AMiBA; Lo et al. 2001),
an interferometric array of 19 elements, will have full polarization
capabilities in order to probe the polarization properties of the CMB.
In this paper, we simulate the AMiBA polarization experiment and study
the topological properties of the AMiBA mock maps through the genus
statistic of $Q$ and $U$ Stokes parameter fields.
We also investigate the effect of polarized Galactic foreground emission
on the CMB polarization fields.

\section{Simulating the AMiBA Polarization Experiment}

\subsection{Observational Strategy}

An efficient observational strategy is essential to image the CMB polarization
field. Although Subrahmanyan (2001) suggests {\it drift scanning} that is
more desirable in reducing the environmental response and the cross-talk,
we adopt simple observational strategy that the co-mounted array can be
rotated and shifted to the adjacent points after some data acquisition.
We assume that the primary beam $A(\bx)$ of an elemental aperture with
diameter 0.3 m is Gaussian, normalized to unity at the peak and with the FWHM 
of $44\min$, the bandwidth of the dual polarization receivers is 20 GHz 
centered at 90 GHz, and the system temperature is 75 K.
When 0.3 m apertures are used, the AMiBA will be sensitive to CMB polarization
over the range $700 < \ell < 2000$ with signal to noise ratio of about 4 at
$\ell \sim 700$ and 2 at $\ell \sim 1150$ in 24 hours (Lo et al. 2001).

We assume that 19 elements are hexagonally close-packed on a platform,
as shown in Figure 1$a$, where the distances between adjacent aperture centers
are 35 cm, giving 171 baselines with the minimum spacing of 35 cm.
This configuration gives a good sampling of the $uv$-plane, and is useful for
reconstructing the CMB polarization maps from the visibility data.

In a single pointing observation, the visibilities are assumed to be measured
for 2 hours at a set of $\bu$ specified by the close-packed configuration,
then the instrument is rotated by $5\deg$ about an axis through the center
of the aperture plane to obtain a different set of $\bu$ with the same baseline
lengths but with different orientations.
In this way the $uv$-plane will be well-covered, allowing for good imaging.
The number of rotations are chosen to be 12 times per pointing for better
sampling of the $uv$-plane.
The $uv$ coverage after a 24 hour observation is shown in Figure 1$b$.
Figures 1$c$ and 1$d$ show the window function in the $\ell$ space and the beam
pattern arising from our simulation, respectively. For large $\ell \gtrsim 60$, 
we can relate spherical harmonic multipole index $\ell$ with a baseline 
coordinate $u$ by $\ell = 2\pi u$ with a good approximation. 

\begin{figure*}
\centerline{\epsfysize=15cm\epsfbox{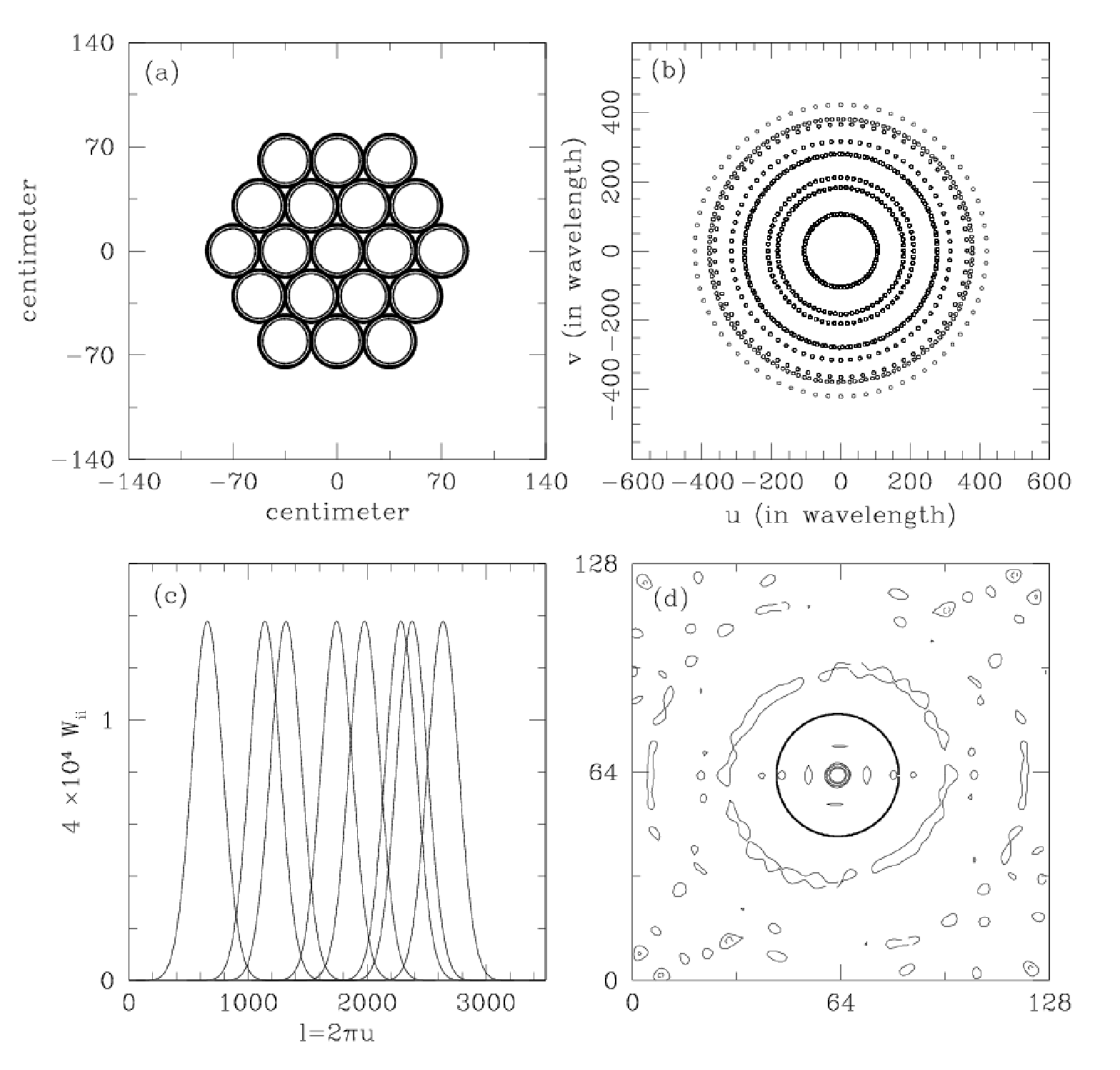}}
\vskip 0.5cm
\hskip 2.9cm {\begin{minipage}{12.5cm}
{\bf Fig. 1.}---~~($a$) A hexagonally close-packed configuration for AMiBA 
         experiment, ($b$) the $uv$ coverage a 24 hour observation,
         ($c$) the window function in the $\ell$ space, and
         ($d$) the beam pattern for a given set of $uv$ coverage
         in a $2\fdg5 \times 2\fdg5$ field. A thick circle in the middle
         shows the FWHM ($44\min$) of the primary beam $A(\bx)$. 
\end{minipage}}
\end{figure*}

The resolution in the $\bu$ space is limited by the area of the sky that
we have surveyed. This is equal to the size of the primary beam in a single 
pointing observation. We can increase the resolution of the map as well as 
the survey area by mosaicking several contiguous pointing observations
(see White et al. 1999 for more details).
Mosaicking does not increase the range of $\bu$. It simply enhances our
resolution by allowing us to follow more periods of a given wave, which is
analogous to the Fraunhofer diffraction through many holes.
In our simulated observation, we perform a mosaic survey of $9 \times 9$ 
pointings with spacing ${\rm FWHM}/2$.

\subsection{Simulated Observations}

The visibility function sampled at a pointing position $\bx_p$ on the sky 
is the Fourier transform of the true sky brightness $I(\bx)$ weighted 
by the primary beam $A(\bx-\bx_p)$, i.e.,
$$
    V({\bu},{\bx_p}) = \int A({\bx-\bx_p}) I({\bx})
             e^{i 2\pi {\bu\cdot\bx}} d^2 {\bx}
    \eqno (1)
$$
(see, e.g., Ng 2001; Hobson, Lasenby, \& Jones 1995).
The size of the primary beam $A(\bx)$ determines the area of the sky that is
viewed and hence the size of the map, while the maximum spacing determines
the resolution.

In order to simulate an observation of CMB polarization fields, we follow
the prescription given in Seljak (1997) and Zaldarriaga \& Seljak (1997).
We use the CMBFAST package (Seljak \& Zaldarriaga 1996; Zaldarriaga, Seljak, 
\& Bertschinger 1998) to get the polarization power spectrum ($C_{\ell}^E$) and
the cross-correlation between temperature and polarization ($C_{\ell}^C$)
in a flat $\Lambda$CDM model with $\Omega_\Lambda = 0.6$, $\Omega_0 = 0.4$,
$h=0.6$, and $\Omega_b h^2 = 0.0125$ (Ratra et al. 1997) in which only the
E-mode (scalar mode) polarization exists and the B-mode vanishes.
To generate small patches of Stokes parameter $Q$ and $U$ fields 
($5\deg \times 5\deg$) on the sky we use small-scale limit approximation 
by Zaldarriaga \& Seljak (1997)
$$
   Q(\bx) = \int d^2 \bu [E(\bu) \cos 2\phi_{\bu}
              - B(\bu) \sin 2\phi_{\bu}] e^{-i 2\pi \bu \cdot \bx},
$$
$$
   U(\bx) = \int d^2 \bu [E(\bu) \sin 2\phi_{\bu}
              + B(\bu) \cos 2\phi_{\bu}] e^{-i 2\pi \bu \cdot \bx},
   \eqno (2)
$$
where $E(\bu)$ is the Fourier component in $\bu$ space of E-polarization
field, $\phi_{\bu}$ is the direction angle of the two-dimensional vector
$\bu$, and $B(\bu) = 0$.
Figure 2$a$ shows a realization of a pure CMB $Q$ field in the $\Lambda$CDM
model in a $5\deg\times 5\deg$ patch on the sky.

\begin{figure*}
\centerline{\epsfysize=19cm\epsfbox{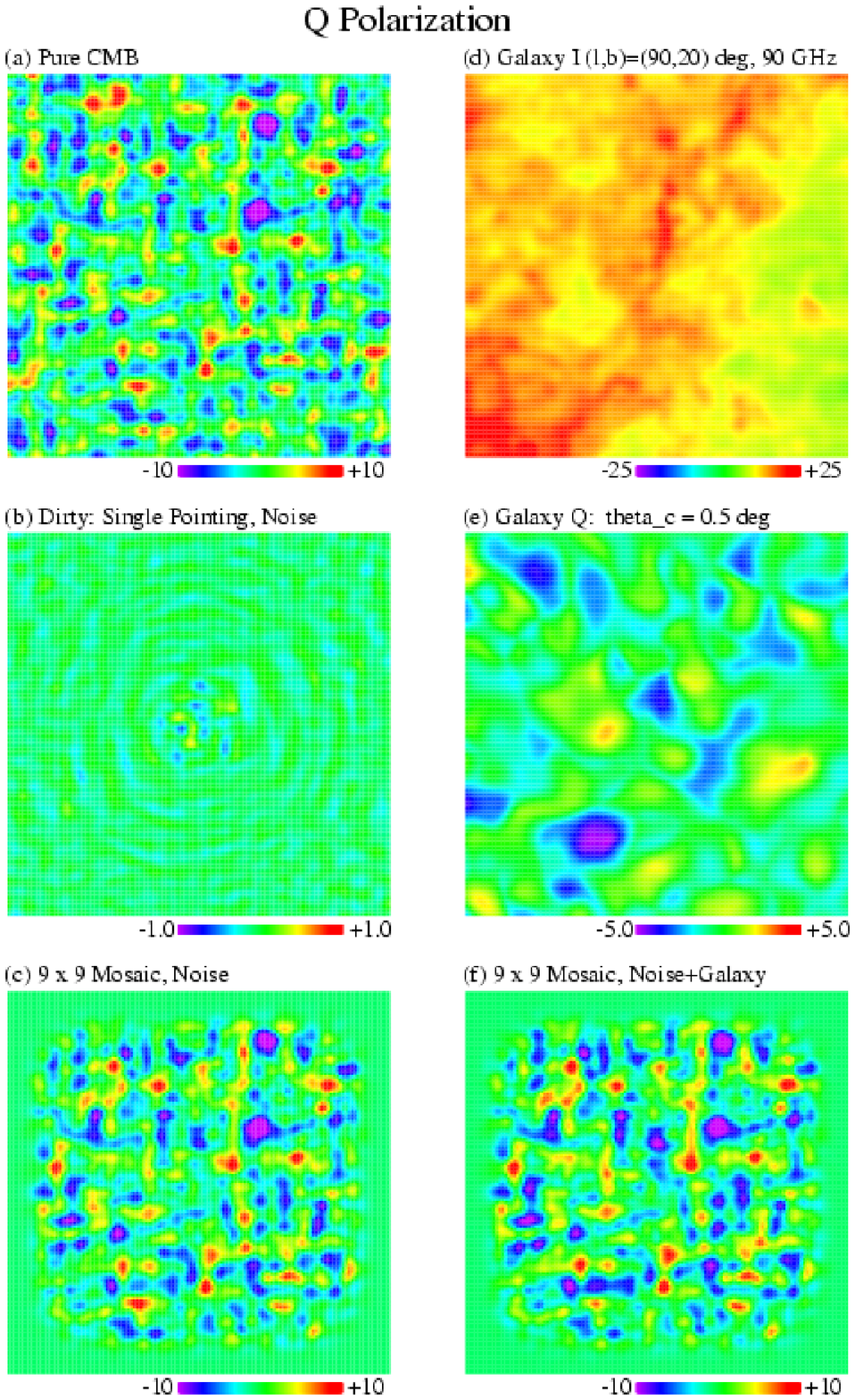}}
\vskip 0.5cm
\hskip 2.9cm {\begin{minipage}{12.5cm}
{\bf Fig. 2.}---~~Examples of mock CMB $Q$ polarization fields, Galactic
            foregrounds, and MEM reconstructed fields. All numbers given
            in colorbars have an unit of $\uK$.
\end{minipage}}
\end{figure*}

According to equation (1), these fields are then multiplied by the primary 
beam and Fourier transformed to give a regular array of visibilities.  
An AMiBA observation is simulated by sampling the regular array 
at the required points in the $uv$-plane specified by the AMiBA observational 
strategy (see Fig. 1$b$).

The AMiBA instrument noise on the data is simulated by adding a random 
complex number to each visibility whose the real and imaginary parts are
drawn from a Gaussian distribution with the variance of the noise predicted 
in a real observation. Here we use the sensitivity per baseline per 
polarization defined as (Wrobel \& Walker 1999; Ng 2001)
$$
   s_b = {1 \over \eta_s\eta_a} {{2k_B T_{sys}} \over {A_{phys}}}
         {1 \over \sqrt{2 \Delta\nu \tau_{acc}}},
   \eqno (3)
$$
where $\Delta\nu$ is bandwidth, $\tau_{acc}$ is the correlator accumulation
time, $A_{phys}$ is the physical area of an elemental aperture, and
$\eta_s$ ($\eta_a$) is system (aperture) efficiency.
Figure 2$b$ shows a dirty image directly FFTed from the single pointing
mock visibility data. Since the dirty image contains biased field information,
it cannot be used for topology analyses. Thus we need to reconstruct
the original and unbiased CMB field from the observed noisy visibility data.

\subsection{Image Reconstruction Using the Maximum Entropy Method}

From the mock visibility data, CMB polarization fields can be reconstructed 
using the maximum entropy method (MEM), in which an extended emission field 
can be more effectively restored compared to the common CLEAN algorithm
(see Narayan \& Nityananda 1986 for a review, and see Conwell \& Evans 1985; 
Chae \& Yun 1994 for applications).
The MEM is also more appropriate for reconstructing a non-Gaussian field, 
compared to the Wiener-filtering method (see, e.g., Maisinger et al. 1998).  
We use the methods of Cornwell \& Evans (1985) for general the MEM algorithm, 
and of Cornwell (1988) for MEM mosaicking.
However, the standard MEM method contains a logarithmic term that is
inapplicable to images that have both positive and negative values,
such as fluctuations in the CMB.
Therefore, we consider the image to be the difference between
two positive additive distributions, namely $h_i = f_i - g_i$, and
the entropy becomes
$$
   S(\bhmap,\bmmap) = \sum_{i} \bigg[\psi_i - 2m_i - h_i\ln\bigg({{\psi_i+h_i}
                      \over {2m_i}}\bigg)\bigg],
   \eqno (4)
$$
where $h_i$ is the predicted map, $m_i$ is
the default model map (Maisinger, Hobson, \& Lasenby 1997; Jones et al. 1998),
and $\psi_i = [h_i^2+4m_i^2]^{1/2}$.

\begin{figure*}
\centerline{\plotfiddle{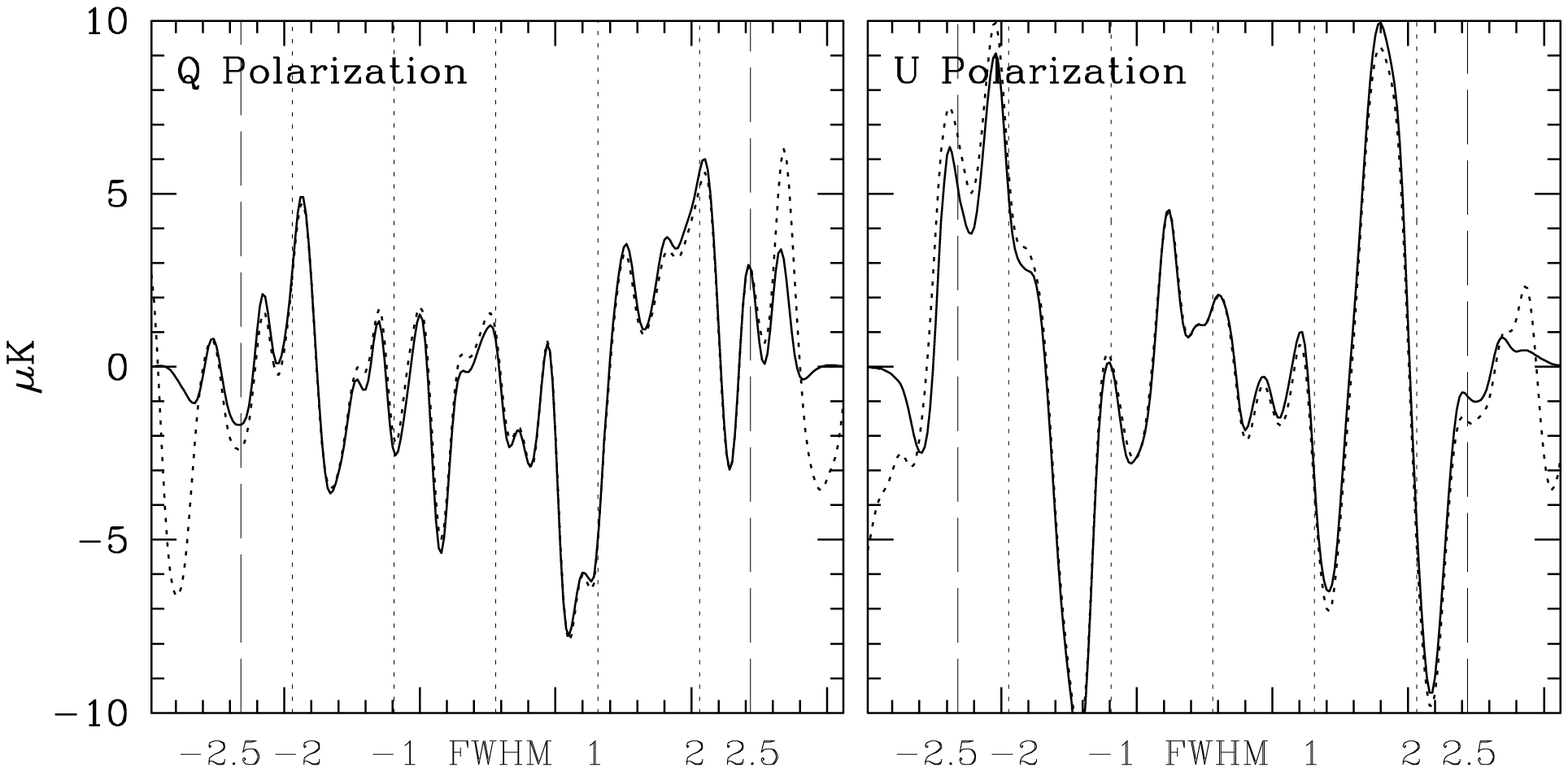}{5 truecm}{0.0}{60}{60}{-440}{-200}}
\vskip 1.2cm
\centerline{ \begin{minipage}{12.5cm}
{\bf Fig. 3.}---~~Profiles of  the CMB $Q$ and $U$ fields along a row 
       in the original maps (dotted curves) and in the MEM reconstructed 
       maps (solid curves). \end{minipage}}
\end{figure*}

The attraction of MEM is that it defines the best image as the solution of
a simple optimization problem obtained by maximizing the entropy
while fitting to the data. Thus extra information can simply be added
as constraints in the optimization.
In our application of mosaicking, $\chi^2$ is defined as
$$
    \chi^2 = \sum_p \sum_r {{|V(\bu_r,\bx_p)
          - \hat V(\bu_r,\bx_p)|^2} \over {\sigma_{V,r,p}^2}}.
    \eqno (5)
$$
Here $\hat V$ denotes the predicted visibility,
and $\sigma_{V,r,p}^2$ is the variance in the visibility of the $r$th
visibility sample at the $p$th pointing (see Cornwell 1988 for more details).

Maximizing
$$
   J \equiv S - \alpha \chi^2,
   \eqno (6)
$$
where $\alpha$ is a Lagrange multiplier, gives a solution for $\bhmap$
that maximizes the entropy and fits to the data. We find the most appropriate
image iteratively using the Newton-Raphson method, i.e.,
$$
    \Delta \bhmap = (-{\nabla}{\nabla} J)^{-1} \cdot \nabla J.
    \eqno (7)
$$
Here we choose the default model map as $\bmmap = 5$ $\uK$, 
constant everywhere, and the Lagrange multiplier as $\alpha = 100$. 
These values are insensitive to the final solution. The mosaicked images 
reconstructed by MEM are very close to the original CMB polarization maps 
(compare Fig. 2$c$ with Fig. 2$a$).
Figure 3 compares the CMB $Q$ and $U$ fields along a row in the original map 
(dotted curves) with those in the MEM map reconstructed from $9 \times 9$
mosaic data (solid curves).

\subsection{Galactic Foregrounds}

We investigate the effects of polarized Galactic foreground emission on the
CMB polarization fields. Two possible sources of the polarized Galactic
emission are synchrotron and spinning dust emissions, which are known to
induce at least 10 \% polarization fraction at frequencies below 90 GHz
(Lubin \& Smoot 1981). Vibrational dust and free-free emissions are expected
to be unpolarized. We use a simple toy model of  Kogut \& Hinshaw (2000) 
to make foreground polarization maps.
We assume here that mean fractional polarization is 10 \%, the polarized
Galactic emission is proportional to the unpolarized intensity, and the effect
of synchrotron emission is negligible at 90 GHz.
We can model the Stokes $Q$ and $U$ components as
$$
   Q = f \cos(2\gamma)I, ~~ U = f \sin(2\gamma)I,
   \eqno (8)
$$
where $f(\btheta)$ is the fractional polarization, assumed to vary across
the sky with $\langle f \rangle =0.1$, and $\gamma(\btheta)$ is the 
polarization angle which is randomly realized with coherence angle 
$\theta_c = 0\fdg5$ (see $\S$2 of Kogut \& Hinshaw 2000 for more details).
We use a two-component dust model of Finkbeiner, Davis, \& Schlegel (1998)
to predict 90 GHz vibrational dust intensity, and a model of Kogut et al.
(1996) for adding the free-free emission. The 100 $\um$ dust map of Schlegel,
Finkbeiner, \& Davis (1998) is used as a basic template map.
Figures 2$d$, 2$e$, and 2$f$ show the 90 GHz Galactic intensity at $b=20\deg$,
its $Q$ polarized contribution, and the MEM reconstructed CMB $Q$ polarized
field contaminated by the Galactic polarized foreground, respectively.

\section{Topology of the Simulated Maps}

The CMB polarization field, like the temperature anisotropy, can be used
for an observational test of the Gaussianity of the primordial density 
fluctuation, and thus provide an important constraint on the inflationary 
models.
We use the two-dimensional genus statistic introduced by Gott et al. (1990)
as a quantitative measure for the topology of the CMB polarization fields.
The genus is defined as the number of hot spots minus the number of cold spots,
or equivalently,
$$
    g(\nu) = {1 \over {2\pi}}\int_C \kappa ds,
   \eqno (9)
$$
where $\kappa$ is the signed curvature of the iso-temperature contours $C$.
Since the genus curve as a function of the temperature threshold level has
a form of $g(\nu)$ $\propto$ $\nu e^{-\nu^2 /2}$ for a Gaussian random-phase 
field, non-Gaussianity of a field can be detected from deviations of the 
genus curve from this relation. For example, non-Gaussianity can cause shift 
or asymmetry of the genus curve.

\begin{figure*}
\centerline{\plotfiddle{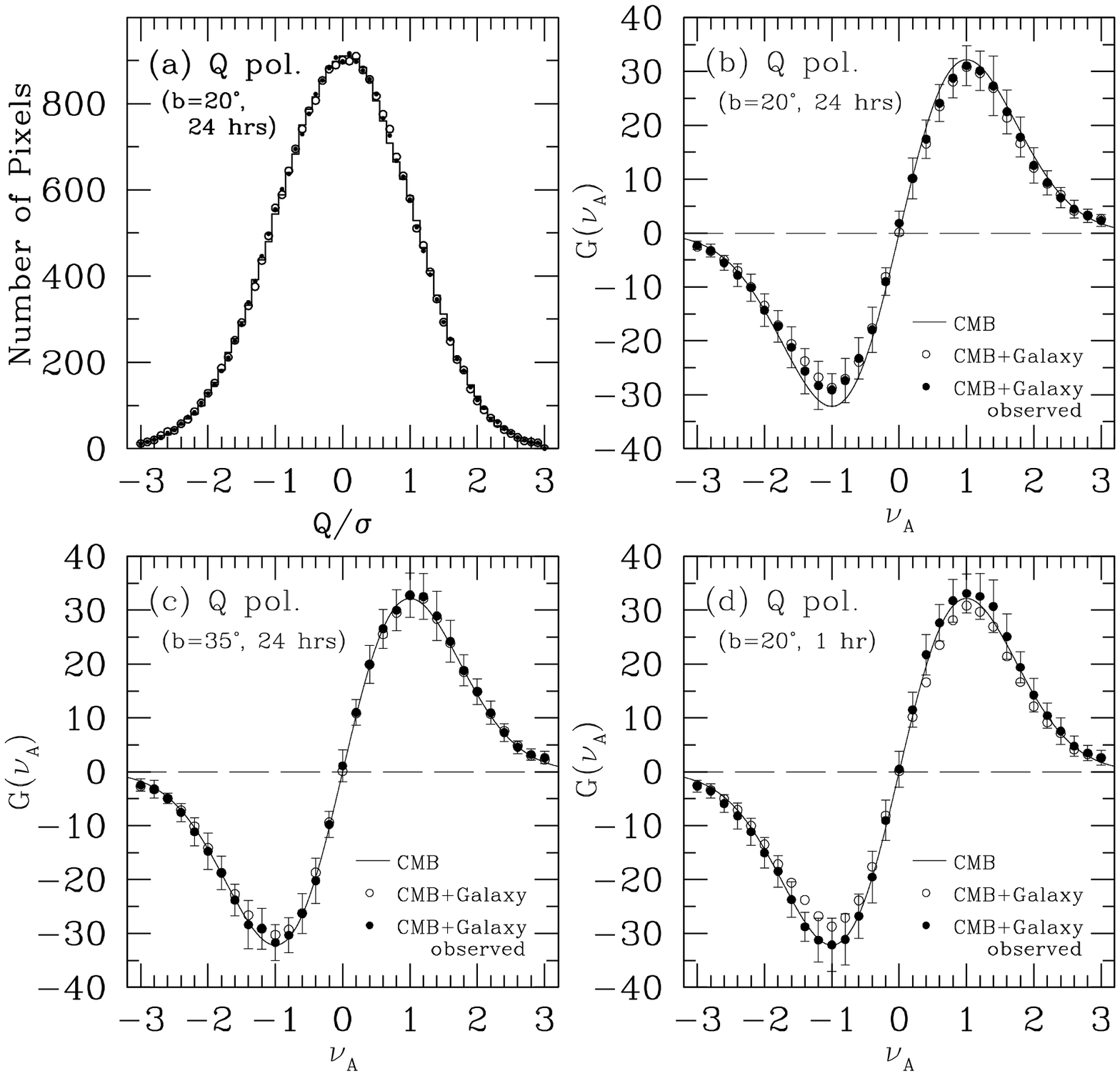}{5.3 truecm}{0.0}{60}{60}{-440}{-260}}
\vskip 5.5cm
\centerline{ \begin{minipage}{12.5cm}
{\bf Fig. 4.}---~~Histograms and genus measured from CMB polarization 
                  $Q$ maps. 
\end{minipage}}
\vskip -0.2cm
\end{figure*}

We present genus curves as a function of the area-fraction threshold level
$\nu_A$, which is defined to be the temperature threshold level at which
the corresponding iso-temperature contours enclose a fraction of the survey
area equal to that of a Gaussian field (Gott et al. 1990).
All measured values are averages over 15 mock AMiBA surveys.
We use the 4 FWHM $\times$ 4 FWHM region in each mock survey field,
where the MEM reconstruction gives more confident results.
Figures 4$a$ and 4$b$ shows averaged histograms and genus for foreground-free
(solid curve), Galactic foreground-added (open circles), and Galactic
foreground-added and MEM reconstructed ($9 \times 9$ mosaic observation, 
filled circles) CMB $Q$ fields. The last case is a result of a 24 hour 
integration per pointing located at Galactic latitude $b=20\deg$.
Note that the solid curve in Figure 4$b$ shows the functional form expected
for a random-phase Gaussian field, $A\nu e^{-\nu^2 /2}$, and has been fitted
to the foreground-free genus result by adjusting $A$.
The result from a 24 hour integration of a region centered at $b=35\deg$
is shown in Figure 4$c$. When the integration time is only one hour per
pointing, we obtain genus curves shown in Figure 4$d$.

Table 1 lists skewness $\langle (Q/\sigma)^3 \rangle$, genus-related 
statistics $A$ (amplitude), $\Delta\nu$ (shift parameter), and $\Delta g$ 
(asymmetry parameter) measured from CMB $Q$ fields; see Peebles (1993) 
for skewness and Park et al. (2001) for definition and measurement of 
genus-related statistics, respectively.

\begin{table*}
\begin{center}
\vskip -0.6cm
{\bf Table 1.}~~Skewness and Genus Statistics Measured from Mock CMB $Q$ Fields\
\vskip 0.2cm
\begin{tabular}{l l c c c c}
\hline\hline
 & $Q$ Pol. & $\langle (Q/\sigma)^3 \rangle$ & $A$ & $\Delta\nu$ & $\Delta g$ \\
\hline
 & CMB & $-0.08\pm0.11$ & $52.5\pm4.4$ & $-0.006\pm0.035$ & $0.04\pm0.10$ \\
\hline
$b=20\deg$ & ~+Galaxy & $-0.06\pm0.16$ & $47.9\pm4.7$ & $-0.005\pm0.047$ & $0.04\pm0.12$ \\
 & ~+Galaxy (observed)$^\dagger$ & $-0.05\pm0.15$ & $48.6\pm4.6$ & $-0.014\pm0.044$ & $0.03\pm0.12$ \\
\hline
$b=35\deg$ & ~+Galaxy & $-0.07\pm0.13$ & $51.6\pm3.9$ & $-0.009\pm0.040$ & $0.05\pm0.11$ \\
 & ~+Galaxy (observed)$^\dagger$ & $-0.06\pm0.12$ & $52.9\pm4.2$ & $-0.008\pm0.037$ & $0.02\pm0.09$ \\
\hline
\end{tabular}
\end{center}
\vskip -0.2cm
~~~~~~~~~~~~~\scriptsize$^\dagger$ $9 \times 9$ mosaic observation with 24 hours
integration time per pointing.
\vskip -0.2cm
\end{table*}

The results show that the polarized Galactic emissions, even at low
latitude, do not have significant effect on the CMB polarization fields.
Although the polarized Galactic foreground emissions at $b=20\deg$ cause slight
asymmetry, and reduce genus amplitude (Fig. 4$b$ and Table 1), these effects
are smaller than the statistical fluctuations due to the sample variance of
the pure CMB map.
The open and filled circles in each panel in Figure 4 show that
the original polarization fields can be well restored from the AMiBA mosaic
observational data by MEM image reconstruction technique if the integration
time is 24 hours per pointing. However, if the integration time is one hour,
instrumental noise increases the amplitude of the genus curve, and the presence
of non-Gaussian signals due to the Galactic emission is buried.

\section{Conclusions}
We have studied the topology of CMB polarization maps by simulating the
AMiBA experiment and by measuring the skewness and genus statistics.
The MEM turns out to be very useful in reconstructing the CMB polarization 
$Q$ and $U$ fields from the interferometric visibility data.
Our study shows that although the polarized Galactic foreground emissions
at low latitude ($b=20\deg$) decrease the amplitude of the genus curve 
and cause asymmetry of the curve, these effects are not significant
compared with the sample variance in our $2\fdg9\times 2\fdg9$ surveys.
At higher Galactic latitudes, the effects of the Galactic emissions i
s shown to be negligible, and the intrinsic non-Gaussianity of the CMB 
polarization is easier to detect.

The major source of statistical variance of the genus statistic is the sample
variance induced by our small survey area (4 FWHM $\times$ 4 FWHM $= 8.6
~{\rm deg.}^2$). This can be reduced not by observing many small different 
patches of the sky, but by increasing the area of the survey at each location.
Therefore, increasing the survey area and integration time at a given pointing
is essential to accurately estimate the effects of Galactic foregrounds 
on the polarization maps and to detect non-Gaussian signals of cosmological 
origin.
 
%%\acknowledgements 
\vskip 0.4cm
We acknowledge valuable discussions with Kin-Wang Ng, Uros Seljak 
and Ravi Subrahmanyan. This work is supported by the BK21 program 
of the Korean Government.

\end{document}